\documentclass[aps,prb,twocolumn,superscriptaddress,amsmath,amsfonts,amssymb,floatfix]{revtex4-1}
\usepackage{graphicx}

\begin{document}

\title{Stoichiometry, structure, and transport in the quasi-one-dimensional metal, Li$_{0.9}$Mo$_6$O$_{17}$}

\author{J.~L.~Cohn}
\affiliation{Department of Physics, University of Miami, Coral Gables, FL 33124}

\author{P.~Boynton}
\altaffiliation [Present address:]{Physics Department, University of California, San Diego, La Jolla, CA 92093-0319.}
\affiliation{Department of Physics, University of Miami, Coral Gables, FL 33124}

\author{J.~S.~Trivi\~no}
\affiliation{Department of Physics, University of Miami, Coral Gables, FL 33124}

\author{J.~Trastoy}
\affiliation{Department of Physics, University of Miami, Coral Gables, FL 33124}

\author{B.~D.~White}
\affiliation{Department of Physics, Montana State University, Bozeman, Montana 59717}

\author{C.~A.~M. dos Santos}
\affiliation{Escola de Engenharia de Lorena - USP, P. O. Box 116, Lorena-SP, 12602-810, Brazil}

\author{J.~J.~Neumeier}
\affiliation{Department of Physics, Montana State University, Bozeman, Montana 59717}

\begin{abstract}
A correlation between lattice parameters, oxygen composition, and the thermoelectric and Hall coefficients is presented for single-crystal
Li$_{0.9}$Mo$_6$O$_{17}$, a quasi-one-dimensional (Q1D) metallic compound.  The possibility that this compound is a compensated metal
is discussed in light of a substantial variability observed in the literature for these transport coefficients.
\end{abstract}


\maketitle
\thispagestyle{empty}\clearpage

\emph{Introduction.} Li$_{0.9}$Mo$_6$O$_{17}$ known as ``lithium purple bronze'' (LiPB), is a low-temperature superconductor ($T_c\approx 2$~K)
first synthesized and studied in the 1980s.\cite{PBReview,oldwork1,BandStructure1} It has attracted interest more recently for its quasi-one dimensionality and Luttinger-liquid
candidacy.\cite{PES2,BandStructure2,PES3,RecentTheory1,RecentTheory2,RecentAllen}  It is distinguished among quasi-one-dimensional (Q1D) compounds by
the absence of a conventional density-wave transition\cite{Optical,NeumeierPRL} (either charge or spin) throughout a broad temperature range,
$T\geq T_c$. An upturn in its resistivity below $T_M\sim 30$~K may be associated with localization, dimensional crossover or the development
of unconventional (e.g., electronically-driven) charge density-wave order.\cite{Optical,NeumeierPRL}

Values reported for the chain-axis (\emph{b}-axis) electrical resistivity of LiPB\cite{oldwork1,Optical,NeumeierMontgomery,XuPRL,Chen,Wakeham,NewHusseyPRL}
vary by more than an order of magnitude, from 0.4 m$\Omega$cm to more than 10 m$\Omega$cm at 300 K, and appear to be responsible for similar differences
in reported values for the anisotropy ratio, $\rho_c/\rho_b$ ($\rho_c$ is the resistivity along the next most conducting direction).  Some
of the differences may be associated with inadequate shorting,\cite{NewHusseyPRL} for current along the chains, of the voltage drops in directions transverse to the current flow,
a delicate matter in highly anisotropic conductors.  Increases in $\rho_b$ for crystals\cite{BoseMetal} annealed in air at 200$^{\circ}$C suggest that
a variable oxygen stoichiometry also contributes to the reported differences. Recent thermoelectric and Hall measurements\cite{LMONernst} support such a view,
indicating small differences in structure and stoichiometry that correlate with transport.
This is an important issue given the renewed attention being paid to LiPB and the sensitivity of conduction in such low-dimensional materials to defects or
impurities that may alter the low-energy electronic structure.
A variable stoichiometry that allows for control of charge-carrier density could prove useful in the study of LiPB which is proposed as a model, bulk system
exhibiting Luttinger-liquid physics.

The principal finding reported here is a correlation between oxygen content and the \emph{c} lattice parameter for a number of as-grown LiPB crystals.
Seebeck (thermopower) and Hall coefficient measurements on LiPB crystals from our own work\cite{LMONernst} and those of
others,\cite{Chen,Wakeham,BoujidaTEP,BoujidaHall} are then presented together. Measured under open circuit conditions,
the thermopower is not prone to the same challenges of transverse shorting that may complicate comparison of
resistivity measurements.  Hall measurements are also less susceptible to such errors given proper averaging with reversed current and magnetic
field directions.  The low-$T$ chain-axis thermopower varies considerably among the crystals measured,\cite{LMONernst}
and correlates with their \emph{c}-axis lattice parameters. Results for the Hall coefficient show considerable variability
but the role of stoichiometry is less definitive as there are not enough data available
on individual crystals for which structure, stoichiometry and transport coefficients have all been measured.
The possibility that LiPB is a compensated metal, motivated by the behavior of the thermoelectric coefficients,\cite{LMONernst} is
examined in light of the Hall data.
\begin{figure}[b]
\includegraphics[width=3.in,clip]{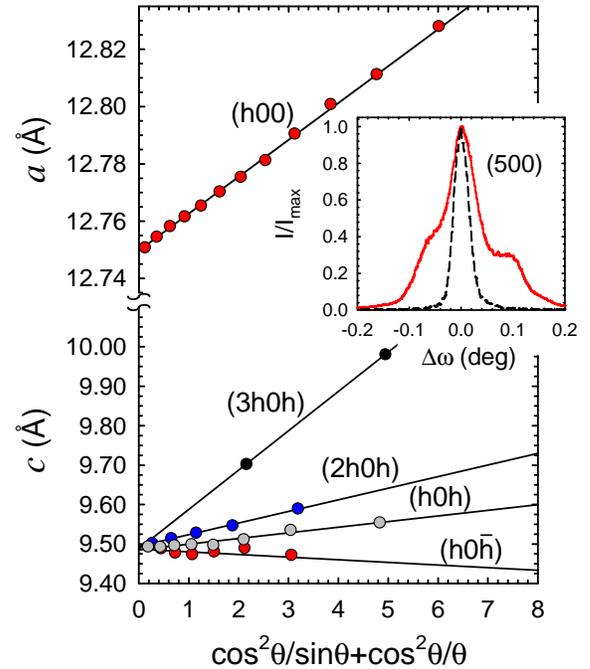}
\caption{(color online) Examples of high-angle extrapolations of XRD reflections used to determine the \emph{a} and \emph{c} lattice parameters for Li$_{0.9}$Mo$_6$O$_{17}$ single crystals.
Solid lines are least-squares fits. Inset: rocking curves of the (500) reflections for two crystals.}
\label{Fig1}
\end{figure}

\emph{Crystal Structure and Stoichiometry.} Single-crystal growth of Li$_{0.9}$Mo$_6$O$_{17}$ using a temperature-gradient flux method is
described in detail elsewhere.\cite{NeumeierPRL,oldwork1} The crystals grow as thin platelets, with $a\simeq 0.05-0.20$~mm and $0.5-2$~mm for \emph{b} and \emph{c}.
A Philips X'Pert x-ray diffractometer (Cu K$_{\alpha}$, $\lambda=1.54056$~\AA) was
employed in determining the crystallographic structure.  Lattice parameters for the monoclinic unit cell\cite{Onoda} were determined by
from high-angle extrapolation\cite{Cullity} of reflections from various lattice planes (Fig.~\ref{Fig1}), and angles $\beta$ (between the $a$ and $c$ axes) from differences
in \emph{d}-spacings for the $(h0h)$ and $(h0\bar{h})$ reflections.  Some crystals exhibit resolution-limited rocking curve widths, FWHM$\ \lesssim 0.05^{\circ}$ (without a monochromator),
while others (typically larger crystals) exhibit a modest mosaicity, FWHM$\ \sim 0.1-0.2^{\circ}$ (inset, Fig.~\ref{Fig1}). For the latter, rocking curves exhibit a convolution of
two or more resolution-limited peaks, indicative of separate domains misaligned by small angles.
A total of ten crystals were studied and their \emph{a} and \emph{b} lattice parameters were found to be the same within
uncertainties, 12.752(2) \AA\ and 5.520(2) \AA, respectively. These agree with those
determined from recent neutron scattering studies\cite{NeumeierNeutron} on a
powder specimen ground from thousands of similarly prepared crystals from our group.

\begin{figure}[t]
\includegraphics[width=3.125in,clip]{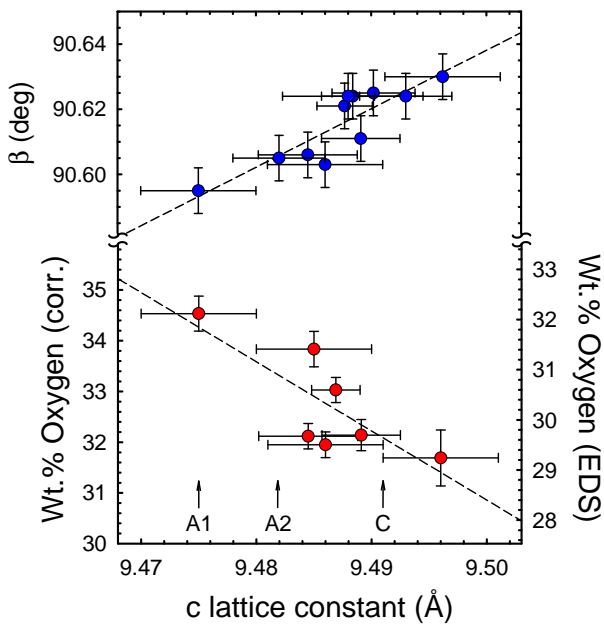}
\caption{(color online) Lower panel: Oxygen content \emph{vs.} \emph{c} lattice constant for Li$_{0.9}$Mo$_6$O$_{17}$ crystals.  Right (left) ordinate scales
correspond to EDS (EDS+2.5 Wt.\%) oxygen contents (see text).  Error bars represent the standard deviation of
EDS oxygen contents from multiple locations and of high-angle $c$-parameter extrapolations from multiple planes.
Arrows labeled A1, A2, and C indicate the \emph{c} lattice parameters for
specimens discussed in the text. Upper panel: $\beta$ \emph{vs.} \emph{c} for all crystals studied.}
\label{Fig2}
\end{figure}
The \emph{c} lattice parameters were found to vary among the as-prepared crystals,
correlating with oxygen content for seven of the crystals that were examined with x-ray energy-dispersive spectroscopy (EDS) on a
system equipped with a polymer window suitable for light-element detection. Such variations, observed in the present study
for crystals from the same growth batch and from different batches, is common in flux-grown crystals.
The results from three or more EDS scans (with area 1 $\mu$m $\times$ 1 $\mu$m) from different locations on the surface
of each crystal were averaged to determine the oxygen content.  Li, not detectable in these scans, was not included in the compositional normalization
[chemical analysis of the neutron scattering powder specimen\cite{NeumeierNeutron} indicated a Li content 0.924(9)].
To ensure that the crystal surfaces (probed to a depth of several $\mu$m by EDS) were representative of the bulk, pieces of two crystals were
polished to half their original thickness (corresponding to a depth of 50-100 $\mu$m) and examined by EDS for comparison.  The average oxygen contents
determined from measurements at several locations on the interior surfaces of these crystals differed from those of the original surfaces by no more
than variations between different locations on each surface.

Fig.~\ref{Fig2} shows the correlation between \emph{c} and oxygen content for these crystals.
Standardless oxygen determinations from EDS typically show small systematic discrepancies due to inadequate accounting for absorption.
As a measure of this discrepancy we use the neutron scattering results\cite{NeumeierNeutron}
which indicate full oxygenation (17 O/f.u., corresponding to 32.09 Wt\% ignoring Li) for $c=9.4909(2)$ \AA.  Comparing our data for the
same \emph{c} values suggests that the EDS oxygen results are $\approx 2.5$~Wt\% too low. The right (left) ordinates of Fig.~\ref{Fig2} correspond
to the EDS (corrected: EDS+2.5 Wt\%) oxygen contents.
%
%

The maximum variation in oxygen content for the crystals examined, $\sim 2$~Wt\%, corresponds to $\sim$ 1 oxygen atom per f.u.
The corrected EDS oxygen scaling in Fig.~\ref{Fig2} implies that some crystals have excess oxygen, presumably accommodated as interstitials.
On the other hand, the fact that four of the crystals have the same measured oxygen content within uncertainties but differing \emph{c} parameters
suggests that other factors may influence \emph{c}, e.g. variations in the Li concentration\cite{oldwork1} or inhomogeneity of the oxygen
distribution (on a scale smaller than the $1~\mu$m). Though prior work\cite{oldwork1} suggests a very narrow range of Li
nonstoichiometry ($0.87-0.93$ per f.u.), small Li concentration variations between crystals may exist and might correlate
with those observed here for oxygen.  The \emph{c} parameter is thus as a surrogate indicator for stoichiometric
variation in oxygen and possibly Li. We note that a crystal with oxygen excess (labeled ``A1'' in Fig.~\ref{Fig2}) was remeasured
after 18 months stored in a dessicator and found to have larger \emph{c} (labeled ``A2'' in Fig.~\ref{Fig2}), suggesting a
tendency of crystals with oxygen excess to slowly lose oxygen or for oxygen to redistribute, via diffusion, over long periods of time.
Small increases in the angle $\beta$ with increasing $c$ parameter are shown in Fig.~\ref{Fig2}.

\begin{figure}[t]
\includegraphics[width=3.125in,clip]{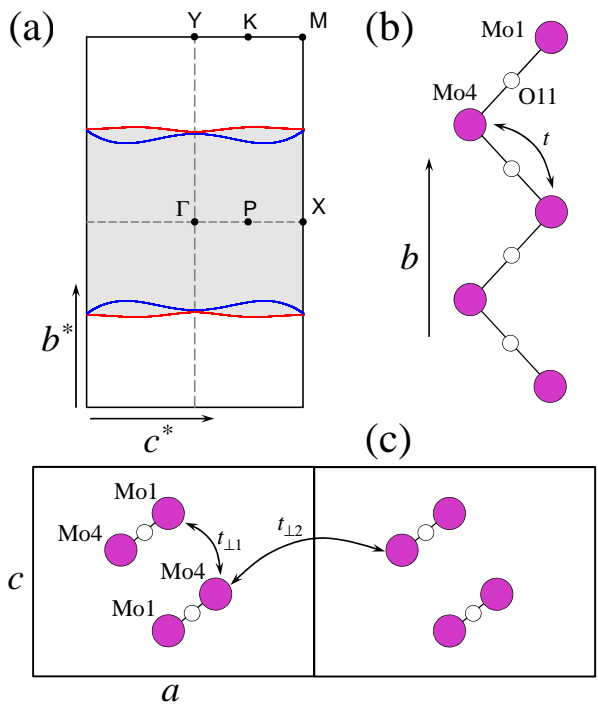}
\caption{(color online) (a) Fermi surface from Ref.~\onlinecite{BandStructure2}, (b) schematic of the conducting Mo-O-Mo zig-zag chains along the crystallographic \emph{b} axis,
and (c) the arrangement of double chains within the \emph{ac} plane.}
\label{FS}
\end{figure}

Interstitial oxygen in other oxides (e.g. cuprates) may affect the electronic structure through local charge doping, lattice distortion, or magnetic moment formation.
The quasi-one-dimensional, conducting Mo-O double chains of LiPB may be especially susceptible to perturbations from such effects.
The LiPB band structure\cite{PES2,BandStructure1,BandStructure2,RecentTheory1,RecentTheory2} is composed of two nearly-degenerate, Q1D bands
crossing the Fermi energy (with $k_F\approx \pi/2b$), yielding two pairs of slightly warped Fermi surface (FS) sheets in the
\emph{b*-c*} planes with very little dispersion along \emph{a*} [Fig.~\ref{FS} (a)]. The small splitting predicted for the bands near $E_F$ from density
functional theory\cite{BandStructure2,RecentTheory2} ($\sim 30-40$~meV) has not been observed in photoemission.\cite{PES2,PES3}  The splitting is associated
with an opposite warping of the FS sheets, attributable
within tight-binding parameterizations\cite{RecentTheory1,RecentTheory2} to opposite signs (and comparable magnitudes) for the hopping integrals between
constituent chains of a ladder ($t_{\perp 1}$) and between ladders ($t_{\perp 2}$) in neighboring unit cells [Fig.~\ref{FS} (b), (c)].  These transverse hopping parameters may be sensitive to
variations in the \emph{c} parameter.
\begin{figure}[t]
\includegraphics[width=3.125in,clip]{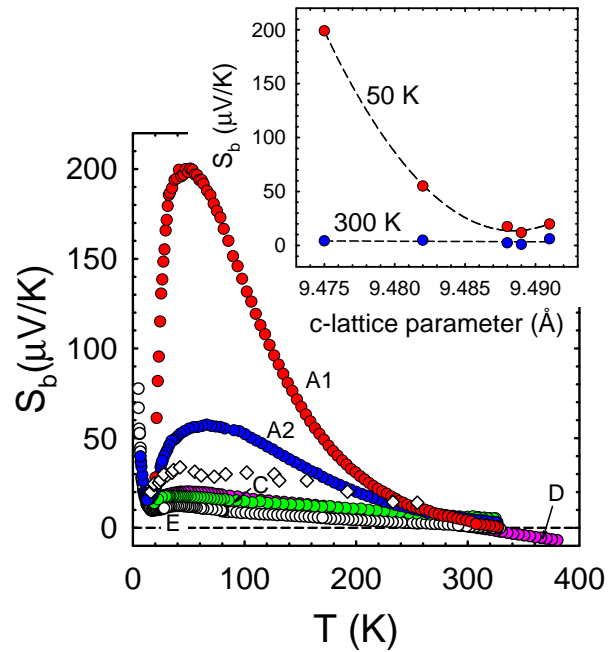}
\caption{(color online) $S_b(T)$ for crystals from Ref.~\onlinecite{LMONernst} (circles with letter labels) and Ref.~\onlinecite{BoujidaTEP} (diamonds). Inset: $S_b$ at $T=$50~K and 300~K
\emph{vs.} \emph{c} lattice parameter for crystals from Ref.~\onlinecite{LMONernst}.}
\label{Fig4}
\end{figure}

\emph{Thermopower.} The \emph{b}-axis thermopower of LiPB, $S_b(T)$ (Fig.~\ref{Fig4}), was first reported in Ref.~\onlinecite{BoujidaTEP}. More recently,\cite{LMONernst} $S_b(T)$ was
reported for several crystals having different \emph{c} parameters (letter labels in Fig.~\ref{Fig4}).  Data for crystal D have been extended in the present work to
385~K.  The inset of Fig.~\ref{Fig4} shows $S_b$ as a function of \emph{c}.  While the room-temperature thermopower is similar for all crystals, the dramatic low-$T$ peak in
$S_b$ near $T=50$~K correlates with the \emph{c} parameter.   The very large value (200 $\mu$V/K) found for specimen A1 with the
smallest \emph{c} parameter and highest oxygen content (Fig.~\ref{Fig2}) is unusual for a metal.  The largest Nernst coefficients ($\nu$), with maxima near $T=20$~K,
were found\cite{LMONernst} for specimens C and D having the smallest values for $S_b$. This favors a two-carrier (ambipolar) conduction scenario in which very large hole and electron
partial thermopowers, $S^h$ and $S^e$ ($S^e<0$), nearly cancel due to compensation: $S_b\propto S^h+S^e$, $\nu\propto \mu (S^h-S^e)$ ($\mu$ is the carrier mobility).
A departure from the compensation condition in specimens with larger \emph{c} parameter could explain the appearance of their large low-$T$ values for $S_b$.
The mechanism underlying such very large partial thermopowers at low $T$ remains to be established, but phonon-drag
appears to be a plausible candidate.\cite{LMONernst}

Though all specimens have positive 300~K thermopowers, their $T$ dependencies approach a linear-in-$T$ form with negative slope that results
in negative values for $S_b$ above room temperature (specimen D, Fig.~\ref{Fig4}). This implies a negative (electron-like) carrier diffusion thermopower along the chains that is
independent of the low-$T$ behavior.

%
\begin{figure}[t]
\includegraphics[width=3.125in,clip]{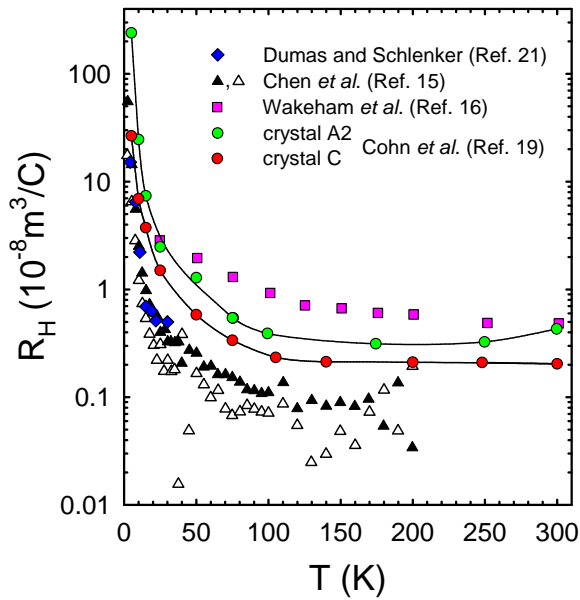}
\caption{(color online) Hall coefficient for crystals A2 and C (Ref.~\onlinecite{LMONernst})
and others from the literature. Data for Ref.~\onlinecite{Wakeham}
were computed by interpolation from their reported
$\sigma_{xy}/B$, $\rho_b$ and $\rho_c$ data as,
$R_H=(\sigma_{xy}/B)\rho_b\rho_c$.} \label{Fig5}
\end{figure}

\emph{Hall Coefficient.} The Hall coefficients of crystals
A2 and C having $c=9.482(4)$ and $9.491(4)$, respectively
(arrows in Fig.~\ref{Fig2}), were reported
elsewhere.\cite{LMONernst}  Their oxygen contents were not
measured; their Hall data are shown in Fig.~\ref{Fig5}. Also shown are
data from three other sources in the
literature.\cite{Chen,BoujidaHall,Wakeham} All of the data show very similar $T$ dependencies, but magnitudes
vary by a factor of 10.  Note that $R_H>0$ throughout the $T$ range for all
measurements.

At present it is not possible to draw conclusions about a dependence of $R_H$ on \emph{c} parameter with complete data sets only for specimens A2 and C
(\emph{c} was not reported along with the other data reproduced in Fig.~\ref{Fig5}).
In a more recent work\cite{NewHusseyPRL} studying crystals from the same source as that of Ref.~\onlinecite{Wakeham}, lattice parameters
$a=12.73$~\AA\ and $c=9.51$~\AA\ where reported, both differing significantly from those reported here.  Thus in addition to
a stoichiometric variability observed for the present crystals, there may well be other
differences between specimens prepared by different groups that have yet to be identified.

The Hall data might be reconciled with the compensation scenario motivated by the thermoelectric data.
Empirically, that $R_H$ is always positive (a sign opposite to that of the diffusion thermopower) and
varies substantially in magnitude among those crystals measured tends to support such a view.
To examine this issue in more detail, first note that
for Q1D systems with open Fermi surfaces the relation between the low-field Hall coefficient and carrier density differs from the free-electron
expression, and for constant mean-free-path is generalized to,\cite{Q1DHall} $R_H=\eta/(n|e|)$, where $\eta$
measures the nonlinearity of the dispersion along the conducting chains, $\eta=-(\hbar k_F/v_F)\partial^2\varepsilon/\partial k_b^2|_{\varepsilon_F}$,
and $v_F=\partial\varepsilon/\partial k_b|_{\varepsilon_F}$.

From the perspective of correlation effects, LiPB is one-quarter filled (approximately half an electron per band crossing $E_F$)
with a tight-binding dispersion,\cite{RecentTheory1,RecentTheory2}
$\epsilon(k_b)=-t\cos(k_bb/2)$. The Hall coefficient corresponding to $k_F=\pi/(2b)$ of LiPB\cite{PES2} would then be,\cite{Q1DHall,Q1DHallTB,Q1DHall2}
$R_H=-(\pi/4)/(n|e|)\simeq -1.7 \times 10^{-9}\ {\rm m}^3{\rm /C}$ assuming 1.9$e^{-}$/unit cell as dictated by the chemistry and bonding.
Though this value is in the middle of the experimental magnitudes near room temperature, the sign is opposite to those observed.
The Hall coefficient for coupled Luttinger chains differs from the noninteracting case only by small power-law-in $T$ corrections.\cite{LLHall}
Theoretical treatments of the thermopower for single Luttinger chains with impurity scattering\cite{LLTEP}
indicate a linear temperature dependence for the carrier diffusion contribution similar to the noninteracting case.

A positive sign for $R_H$ can arise from negative curvature in the dispersion ($\eta>0$)
along one or both of the FS sheets (e.g. the density functional
bands\cite{BandStructure2,RecentTheory2} suggest this is the case for the band that
crosses at higher momentum along the $P-K$ direction), or from a strong momentum dependence for the mean-free path\cite{OngGeometricHall}
due to electron-electron\cite{Q1DHall,ElElmfp} or electron-phonon interactions, though such scattering anisotropy tends to be weaker at high $T$.
Two-carrier fits\cite{LMONernst} to the $R_H$ data for crystals A2 and C yield hole and electron partial Hall coefficients that are
an order of magnitude or more larger than $R_H$, implying carrier densities substantially smaller than expected.
A reduced mobile carrier density is suggested by observations of localized charge in optical studies,\cite{Optical} as might be
anticipated for carriers confined to chain fragments isolated by disorder.

An alternate, intriguing possibility is that strong correlations modify the FS from that predicted within density functional theory [Fig.~\ref{FS} (a)].  The Mott gap ($\Delta$)
in LiPB appears to be quite small\cite{RecentTheory2} such that $t_{\perp}>\Delta$.
In such systems having strong on-site and longer-range Coulomb repulsion, calculations reveal\cite{TsvelikRPA,GiamarchiFermiArcs} a regime intermediate between 1D insulator and 2D metal where
the suppression of the Mott gap via transverse hopping leads to the formation of a FS broken into narrow electron and hole pockets. This physics was recently invoked as a possible
explanation for Fermi arcs found in photoemission studies of Y-124 due to the CuO chains.\cite{Y-124}  Such a picture is appealing for LiPB as it offers a possible explanation
for two-carrier physics apparent in the thermoelectric coefficients, positive values for $R_H$ (provided there is a higher weighting of hole-like pockets),
and carrier densities inferred from a two-carrier analysis of $R_H$ that are much smaller than expected.  Whether this scenario can be consistent with
photoemission observations of the FS awaits a more complete and higher-resolution mapping throughout the Brillouin zone.\cite{RecentAllen}

\emph{Summary.} Variations in the \emph{c}-lattice parameter of Li$_{0.9}$Mo$_6$O$_{17}$ crystals, attributable to stoichiometric variations
in oxygen and possibly Li, correlate with the low-$T$ thermopower.  The implication is that the low-energy electronic structure is sensitive to
small variations in stoichiometry in as-grown crystals.  Thus the \emph{c} parameter is revealed by this work to be an important metric with which to
compare specimens. The data suggest formation of interstitial oxygen during growth as a possible source of oxygen variability.  The available Hall
data are less conclusive as to trends with \emph{c}, but the observed variations in magnitude may be consistent with a two-carrier picture motivated by
thermoelectric data.

The authors acknowledge assistance from Mr. Tom Beasley  and the Florida Center for Analytical Microscopy at Florida International Univ.
This material is based upon work supported by the National Science
Foundation under grant DMR-0907036 (Mont.~St.~Univ.), the Research Corporation and U.S. Department of Energy (DOE)/Basic Energy Sciences (BES)
Grant No. DE-FG02-12ER46888 (Univ.~Miami), and in Lorena by the CNPq (301334/2007-2) and FAPESP (2009/14524-6).

\end{document}